\documentclass{eas}
\usepackage{graphicx}
\newcommand\langeditorchanges[1]{#1}

\newcommand\edchange[1]{#1}
\newcommand\ednote[1]{} 
\begin{document}

\title{Do we expect to find the Super-Earths close to the \langeditorchanges{gas giants?}}
\runningtitle{\edchange{Podlewska E.: The Super-Earths close to the gas giants?}} 
\author{E. Podlewska}\address{CASA* and Institute of Physics, University 
of Szczecin, Poland}
\begin{abstract}
We  have investigated the evolution of a pair of interacting planets
embedded in a gaseous \langeditorchanges{disc,} considering \langeditorchanges{the} possibility of the resonant 
capture of a Super-Earth by a Jupiter mass gas giant.
First, we have examined the situation where the Super-Earth is on
the internal orbit and the gas giant on the external one.
It has been found that the terrestrial planet is scattered from the 
disc or the gas giant captures the Super-Earth into an interior 3:2 
or 4:3 \langeditorchanges{mean-motion} resonance. The stability of \langeditorchanges{the latter configurations} 
depends on the initial planet positions and \langeditorchanges{on} eccentricity
evolution. 
The behaviour of the system is different if the Super-Earth is the
external planet. We have found that instead of being captured in the 
\langeditorchanges{mean-motion} \langeditorchanges{resonance,} the terrestrial planet is trapped at the outer 
edge of the gap opened by the gas giant. \langeditorchanges{This effect}
prevents \langeditorchanges{the} occurrence of the first order \langeditorchanges{mean-motion}
commensurability.
\langeditorchanges{These} results are particularly interesting in light \langeditorchanges{of} recent 
exoplanet discoveries and provide predictions of what will become
observationally testable in the near future. 
\end{abstract}
\maketitle
\section{Introduction}
So far we know only \langeditorchanges{a} few Super-Earths, \langeditorchanges{i.e.,} planets less massive 
than $10 M_\oplus$. However, there is a good chance that in the
near future, COROT and \langeditorchanges{the} Kepler mission
 will find more terrestrial type planets.
An interesting possibility is the existence \langeditorchanges{of} Super-Earths
close \langeditorchanges{to} Jupiter-like planets.
\langeditorchanges{Different} mass objects embedded in a protoplanetary disc will migrate
with different rates. The final configurations will
depend on the intricate interplay among many physical processes including
planet-planet, disc-planet and planet-star interactions.
Here we focus on the possible resonant configurations of a gas giant
and a Super-Earth orbiting a Solar-type star.

The early divergent conclusions concerned with 
the occurrence 
\langeditorchanges{of} terrestrial planets \langeditorchanges{in} hot Jupiter systems (Armitage,
\cite{armitage}, Raymond et al., \cite{raymond05})
have been clarified by the most recent studies (Fogg \& Nelson,
\cite{fonel07a}, \cite{fonel07b}, Raymond et al., \cite{raymond}),
which predict that terrestrial planets can grow and be retained \langeditorchanges{in}
hot-Jupiter \langeditorchanges{systems,} both interior and exterior to the gas giant. 
Here we consider the evolution of an already formed Super-Earth
and a gas giant, both embedded in a gaseous disc. In our studies 
we are interested in the possibility of the 
 formation of resonant planetary configurations.
First, we have examined the behaviour of the system when the terrestrial 
planet is on the internal orbit and the gas giant on the external one. 
Then we have reversed the situation such that the Jupiter is the
\langeditorchanges{internal}
planet and the Super-Earth is \langeditorchanges{the external planet}. 

\langeditorchanges{Two-dimensional} hydrodynamical simulations have been employed in order to determine 
the occurrence \langeditorchanges{of first-order mean-motion} resonances as the outcome \langeditorchanges{of}
convergent orbital migration. Our predictions have \langeditorchanges{interesting}
astrobiological \langeditorchanges{implications,} which we discuss in \langeditorchanges{the Conclusions section}. 

\section{Results of the simulations}
The evolution of the Super-Earth with \langeditorchanges{a mass of} 5.5 Earth masses and the 
Jupiter mass planet embedded in a gaseous disc has been calculated using 
the hydrodynamical code NIRVANA (details of the numerical 
scheme and code can be found in Nelson et al. (\cite{nelson2000})).

The surface density profile at the position of the planets is taken 
to be flat as in 
Papaloizou \& Szuszkiewicz (\cite{papszusz}).
Our computational domain extends between 0.33 and 4 in units normalized in 
such a way that $r=1$ corresponds to  the mean distance of the Jupiter from 
the Sun in the Solar System (5.2 AU). The disc is divided into 
$384 \times 512$ grid-cells in the radial and azimuthal directions, 
respectively.
The radial boundary conditions are taken to be open and the gravitational 
potential is softened with the softening parameter $\varepsilon =0.8H$ where
$H$ denotes the disc semi-thickness.

\subsection{The Super-Earth on the internal orbit}
\begin{figure}
\centerline{
\hbox{
\hbox{\includegraphics[width=12.0cm]{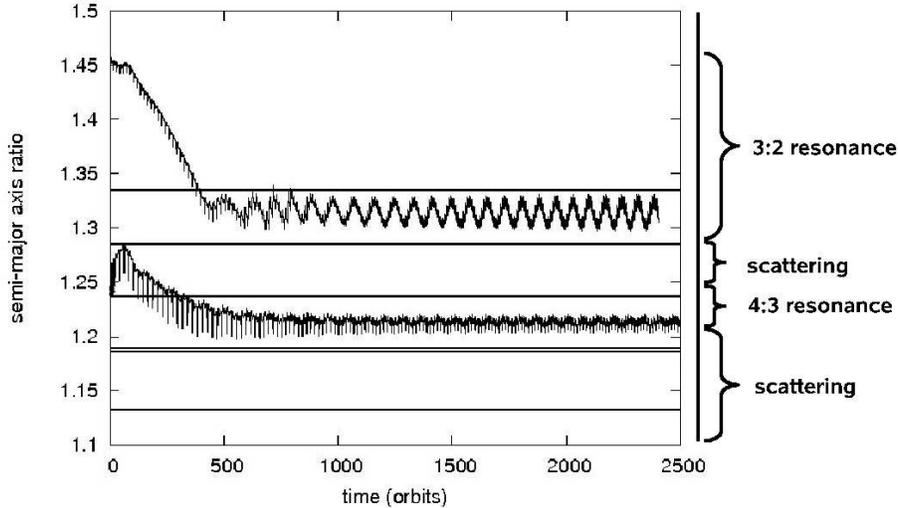}}
}
}
\caption{\label{all}{The evolution of the ratio of semi-major axes 
for the Jupiter and the Super-Earth
embedded in a gaseous disc. In the case 
of \langeditorchanges{the} upper \langeditorchanges{curve,}
the initial separation of the planets is 0.45 and \langeditorchanges{the} 
planets became locked into \langeditorchanges{a}
3:2 resonance. For the lower curve, the initial  planet separation
 is
0.24 and the attained commensurability is 4:3. The solid horizontal lines
are the 3:2, 4:3 and 5:4 resonance widths in the circular three-body
problem. Note that 4:3 and 5:4 resonances partially overlap each other.
}}
\end{figure}

In this part we investigate the evolution of the Super-Earth on the internal
orbit and the Jupiter on the external one. The planets are embedded in the 
flat part of the surface density profile and their relative separation \langeditorchanges{varies}
in the range from 0.15 to 0.45.
In order to \langeditorchanges{get} convergent migration of the planets we have performed the
simulations with \langeditorchanges{a} constant aspect ratio $h$=0.05, \langeditorchanges{a} constant kinematic
viscosity \langeditorchanges{$\nu=10^{-5}$,} which at the initial interior planet's radius 
corresponds to the viscosity parameter \langeditorchanges{$\alpha=4 \cdot 10^{-3}$,} and 
\langeditorchanges{$\Sigma=6 \cdot 10^{-4}$,} which is two Jupiter masses
spread out within the mean distance of the Jupiter from the Sun.
For this set of disc parameters and masses of the planets the Jupiter
migrates faster than  the Super-Earth, and thus the planets approach each 
other. As the evolution proceeds two outcomes are possible, either the
Super-Earth is ejected from the disc or it becomes locked in the 3:2 or 
4:3 mean motion resonances (Podlewska \& Szuszkiewicz, \cite{podszusz}).
The resonances located closer to the Jupiter \langeditorchanges{such as} 5:4 \langeditorchanges{or} 6:5 are
not possible, because for such small separations the system becomes unstable
and the Super-Earth is scattered from the disc. All \langeditorchanges{the} simulations are 
summarized in Fig. \ref{all}.
Moreover, not all planetary pairs initially locked in \langeditorchanges{a mean-motion}
resonance survived \langeditorchanges{further} evolution. In the case of 3:2 commensurability
the libration width increases with time. In fact, in the 
case \langeditorchanges{of}
relative separations 0.26 and \langeditorchanges{0.27,} the amplitude of the oscillation grows 
and eventually the Super-Earth is scattered from the disc. The direct reason 
of this scattering is the excitation of the Super-Earth eccentricity.

\ednote{Please consider to provide figures with better resolution, e.g.,
original Encapsulated Postscript or bitmap files with the resolution of
300dpi.}

\begin{figure}
\centerline{
\hbox{
\hbox{\includegraphics[width=10.0cm]{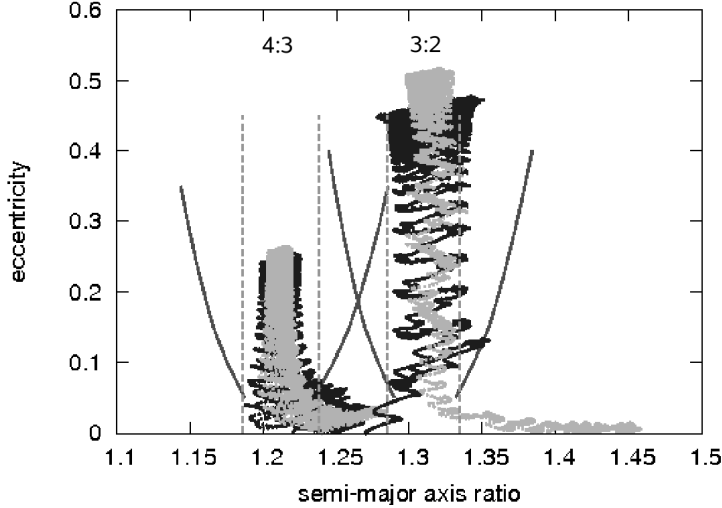}}
}
}
\caption{\label{aaee2}{Four examples \langeditorchanges{of} Super-Earth eccentricity
evolution.
For the 4:3 commensurability we \langeditorchanges{present} the eccentricity
\langeditorchanges{of} the planets \langeditorchanges{with} initial
separation 0.24 (gray colour) and 0.22 (black colour) and for the 3:2 commensurability
for the initial planet separation 0.45 (gray colour) and 0.27
(black colour).
}}
\end{figure}

The eccentricity evolution of the terrestrial planets which survived in 
\langeditorchanges{resonant} configurations till the end of our calculations are shown in 
Fig.2. In this \langeditorchanges{figure,} we plot the eccentricity versus \langeditorchanges{the} semi-major axis \langeditorchanges{ratio,} 
which was calculated by dividing the semi-major axis of \langeditorchanges{the} Jupiter by the 
semi-major axis of the Super-Earth.
For each commensurability the two solid lines show the maximum libration width
obtained from the formula given by Winter \& Murray (\cite{winmur}) in the 
pendulum approximation of the classical restricted
three-body problem.  
The two dashed vertical lines in Fig. 2 denote the resonance
width for the circular case. The eccentricity of the Super-Earth locked in 
3:2 commensurability increases and at the end of simulations reaches a value 
of about 0.5. If the Super-Earth is locked in 4:3 mean motion resonance its 
eccentricity in all simulations approaches to a value around \langeditorchanges{0.3.}

\subsection{The Super-Earth on the external orbit}

\begin{figure*}
\hbox{
\hbox{\includegraphics[width=6.2cm]{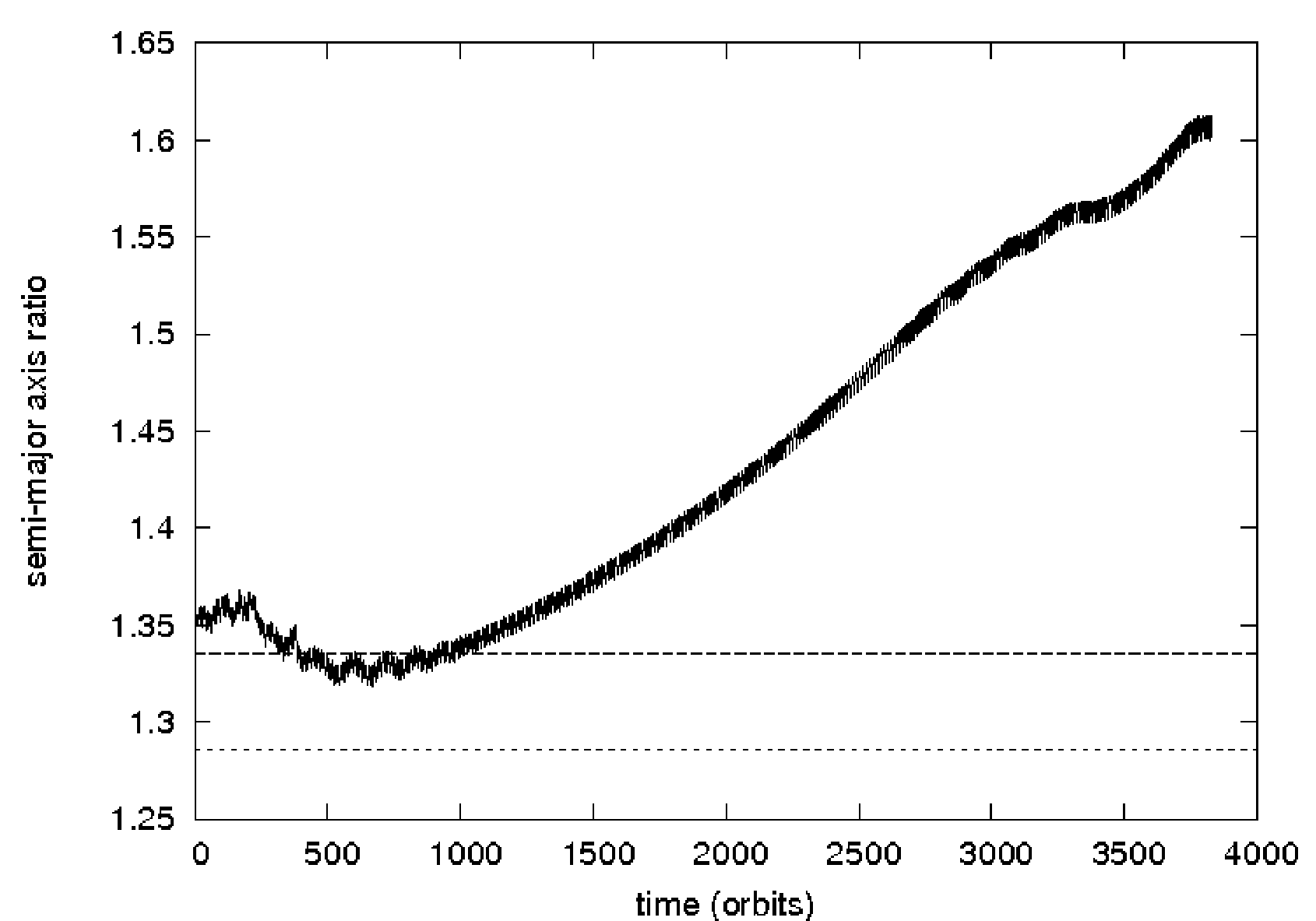}}
\hbox{\includegraphics[width=6.2cm]{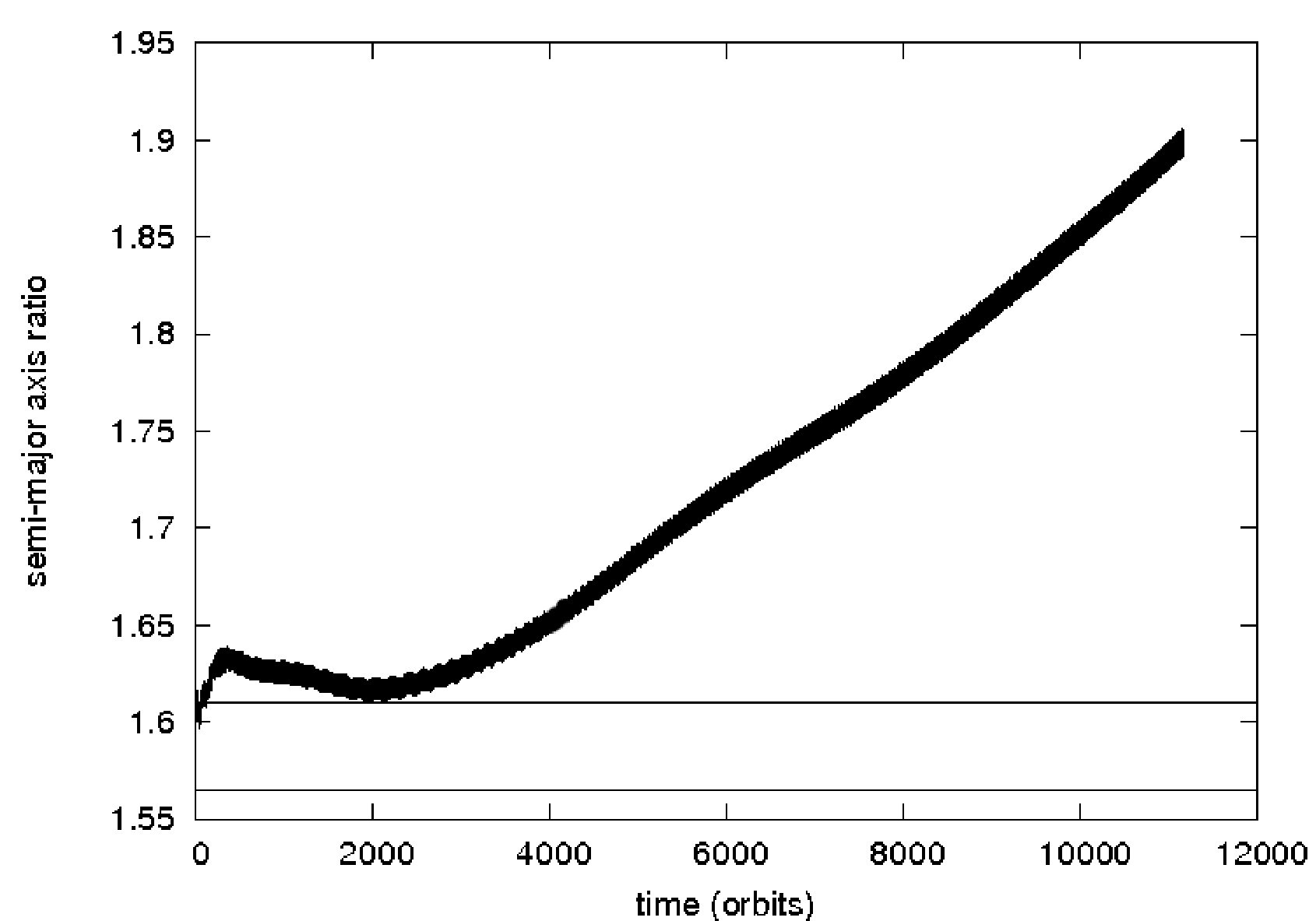}}
}
\caption{\label{fig22}{Left panel: 
The semi-major axis ratio of the planets versus time.
The initial planet separation is 0.35. 
The planets have
been captured in the 2:3 mean-motion resonance but the resonant structure 
did not hold and
the migration became divergent.
The horizontal lines denote the width
of the 2:3 commensurability.
Right panel: \langeditorchanges{As} for the left \langeditorchanges{panel, for} initial
separation 0.62. The horizontal lines denote the width
of the 1:2 resonance.
}}
\end{figure*}

As we have shown in the previous \langeditorchanges{section} it is very likely that the 
Super-Earth on the internal
orbit and the gas giant on the external one form \langeditorchanges{a} resonant configuration. 
Now we ask ourselves if the \langeditorchanges{mean-motion} commensurabilities can
be established when the terrestrial planet is on the external orbit and the
Jupiter is \langeditorchanges{the} internal planet.
\langeditorchanges{By changing} the disc parameters we have succeeded in bringing \langeditorchanges{the} planets close to
each other. \langeditorchanges{Convergent} migration has been achieved for the kinematic
viscosity $\nu=2 \cdot 10^{-6}$, the surface density $\Sigma=6 \cdot 10^{-4}$
and the aspect ratio $h=0.03$.
We have  located the gas giant initially \langeditorchanges{at}
distance 1
 and the Super-Earth further from the star. 
In Fig.~\ref{fig22}a (left panel) we show the evolution of the semi-major axis 
ratio of the planets 
 for the initial separation 0.35. 
The horizontal lines show the width of the 2:3 resonance
(Lecar et al., \cite{lecar}). 
However, the resonant configuration  
lasts just for a few \langeditorchanges{hundred} orbits and after 
that the relative distance between \langeditorchanges{the} planets increases.
Before revealing the reason for which the 2:3 mean motion resonance
cannot be \langeditorchanges{sustained,} let us check the possibility
of the occurrence of the another first order commensurability, namely 1:2.
To this aim, we have placed the Super-Earth further out from the Jupiter,
at  the relative separation 0.62.   
The evolution of this configuration
is shown in Fig.~\ref{fig22}b (right panel). The convergent migration continues
for about 2000 orbits and after \langeditorchanges{that,} similarly \langeditorchanges{to} 
the previous \langeditorchanges{case,} the
migration becomes divergent.

\begin{figure}
\centerline{
\hbox{
\hbox{\includegraphics[width=8.0cm]{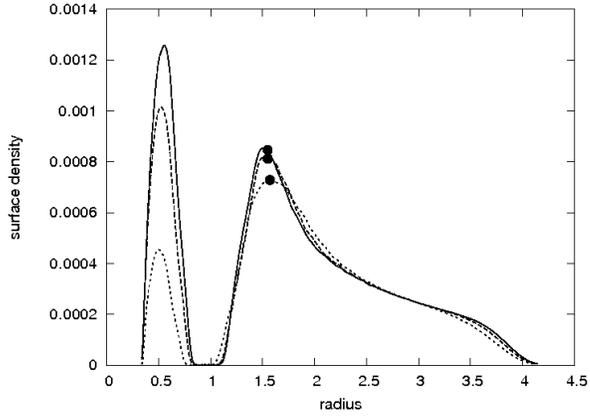}}
}
}
\caption{\label{fig6}{
The surface density profile of the disc after 2355
(solid line), 3925 (dashed line) and 7850 (dotted line) orbits. The dots
denote the positions of the Super-Earth.
}}
\end{figure}

\langeditorchanges{Mean-motion} commensurability cannot be attained because the migration of   
the terrestrial planet is stopped at the outer edge of the gap and this
 is the reason for which the relative distance between \langeditorchanges{the} planets
increases.
The whole evolution can be described as follows. 
The Super-Earth embedded in the disc migrates inward 
toward the Jupiter according to type I migration until it
reaches the outer edge of the \langeditorchanges{gap,}
 where it is trapped. 
\langeditorchanges{This} is illustrated 
in Fig.~\ref{fig6} where 
we show the surface density profile changes
after about 2355, 3925 and 7850 
orbits, together with the position of the terrestrial planet marked
by the black dot. Once the terrestrial planet \langeditorchanges{reaches} the trap it remains
captured till the end of the simulation. Such planetary trapping 
 has been already noticed by Masset 
et al. (\cite{masset}) and Pierens \& Nelson (\cite{pierens}).

\begin{figure}
\centerline{
\hbox{
\hbox{\includegraphics[width=8.0cm]{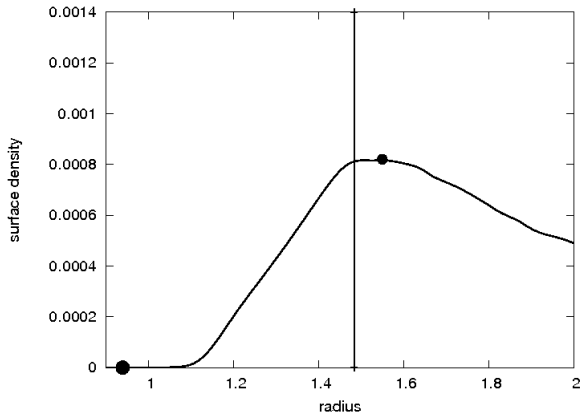}}
}
}
\caption{\label{resgap}{
The surface density profile of the outer edge of the
gap after 3925 orbits. The vertical line denotes the position of the 1:2
commensurability. The dots denote the position of the planets. The
Super-Earth
is captured in the trap and it cannot approach to the resonance.
}}
\end{figure}

The conclusion is \langeditorchanges{that} \langeditorchanges{first-order mean-motion} commensurability cannot be achieved because
the gap is very
wide and the Super-Earth reaches the trap before it is
captured in the
 resonance
(Podlewska \& Szuszkiewicz, \cite{paperII}). 
We have presented this in  Fig.~\ref{resgap} where we plot the gap profile
of the disc after 3925 orbits. 
The dots \langeditorchanges{denote} the position of the Super-Earth and the Jupiter. The vertical
line shows the location of the 1:2 commensurability.
   
\section{Conclusions}

A large-scale orbital migration \langeditorchanges{in} young planetary systems might play an
important role in \langeditorchanges{shaping} their architectures. \langeditorchanges{Tidal} gravitational
forces are able to rearrange the planet positions according to their masses and
the disc parameters. The final configuration after \langeditorchanges{disc} dispersal might
be what we actually observe in the extrasolar systems.
In particular, \langeditorchanges{convergent} migration can bring planets into \langeditorchanges{mean-motion}
 resonances \langeditorchanges{as} has been found in Podlewska \& Szuszkiewicz 
(\cite{podszusz}) \langeditorchanges{in the} case where the
terrestrial planet is on the internal orbit. 

When the Super-Earth is on the
external \langeditorchanges{orbit,} it is
 captured at the outer edge of the gap opened by the gas giant (Podlewska \&
Szuszkiewicz, \cite{paperII}).
Despite \langeditorchanges{the} absence of \langeditorchanges{mean-motion} resonance, the trapping at the
outer
edge of the gap can slow down the migration of the \langeditorchanges{Super-Earth.}
Thus, depending on whether the Super-Earth is inside or outside \langeditorchanges{the} gas
giant \langeditorchanges{orbit,} the most likely planet configurations will be different.
We claim that the Super-Earth can survive in close proximity to the gas
giant. \langeditorchanges{The} terrestrial planet is \langeditorchanges{either} locked
in \langeditorchanges{mean-motion} commensurability or it is captured in the trap which
prevents \langeditorchanges{fast} migration toward the star.

Our results have an interesting implication \langeditorchanges{for} astrobiological studies. 
Namely,
if we have a gas giant in or near the habitable \langeditorchanges{zone,} 
\langeditorchanges{then a} terrestrial planet
locked in the mean motion resonance or captured in the trap at the outer
edge of the gap  can be also located in the habitable zone.
Some of the known extrasolar gas giants as well as the
recently
discovered Super-Earth in the Gliese 581 system 
(Udry et al., \cite{udry}) are in the
habitable \langeditorchanges{zone.} 
 We can expect that future observations will reveal terrestrial planets close
to gas giants.

\section{Acknowledgments}
This work has been partially supported by MNiSW grant N203 026 32/3831
(2007-2010) and  MNiSW PMN grant - ASTROSIM-PL ''Computational Astrophysics.
The formation  and evolution of structures in the universe:from planets to
galaxies'' (2008-2010)
The simulations reported here were performed using the
Polish National Cluster of Linux Systems (CLUSTERIX) and the computational
 cluster "HAL9000"
 of the Faculty of Mathematics and Physics at the University of Szczecin.
I am grateful to Ewa Szuszkiewicz for enlightening discussions and
suggestions
 and Adam {\L}acny for his helpful comments.



\begin{thebibliography}{99}
\bibitem[2003]{armitage}
Armitage, P., 2003, ApJ, 582, L47
\bibitem[2007a]{fonel07a}
Fogg, M. J., Nelson, R. P., 2007a, A\&A, 461, 1195
\bibitem[2007b]{fonel07b}
Fogg, M. J., Nelson, R. P., 2007b, A\&A, 472, 1003
\bibitem[2001]{lecar}
Lecar, M., Franklin, F.  A., Holman, M. J., Murray, N. J., 2001, ARA\&A 39,
581
\bibitem[2006]{masset}
Masset, F. S., Morbidelli, A., Crida, A., Ferreira, J., 2006, ApJ, 642, 478
\bibitem[2000]{nelson2000}
Nelson R. P.\etal\, 2000, MNRAS,
318, 18
\bibitem[2005]{papszusz}
Papaloizou, J. C. B., Szuszkiewicz, E., 2005, MNRAS,  363, 153
\bibitem[2008]{pierens}
Pierens, A., Nelson, R. P., 2008, A\&A, 482, 333
\bibitem[2008]{podszusz}
Podlewska, E., Szuszkiewicz, E., 2008, MNRAS, 386, 1347
\bibitem[2009]{paperII}
Podlewska, E., Szuszkiewicz, E., 2009, MNRAS, 397, 1995
\bibitem[2005]{raymond05}
Raymond, S. N. \etal\ , 2005, Icarus, 177, 256
\bibitem[2006]{raymond}
Raymond, S. N., Mandell, A. M., Sigurdsson, S., 2006, Science, 313, 1413
\bibitem[2007]{udry}
Udry, S. \etal\ , 2007, A\&A, 469, L43 
\bibitem[1997]{winmur}
Winter, O. C., Murray, C. D., 1997, A\&A, 319, 290
\end{thebibliography}
\end{document}